\documentclass[reprint]{revtex4-1}

\usepackage{graphicx}
\usepackage{amssymb}
\usepackage{amsmath}
\usepackage{epsfig}
\usepackage{chemarr}
\usepackage{url}
\usepackage{natbib}

\usepackage{dcolumn}
\usepackage{bm}
\usepackage{color}

\usepackage{comment}

\begin{document}

\title{The difficulty of folding self-folding origami}

\author{Menachem Stern, Matthew Pinson, Arvind Murugan}
\affiliation{Department of Physics and the James Franck Institute,\\ University of Chicago, Chicago, IL 60637} 

\begin{abstract}
\end{abstract}

\begin{abstract}
Why is it difficult to refold a previously folded sheet of paper? We show that even crease patterns with only one designed folding motion inevitably contain an exponential number of `distractor' folding branches accessible from a bifurcation at the flat state. 
Consequently, refolding a sheet requires finding the ground state in a glassy energy landscape with an exponential number of other attractors of higher energy, much like in models of protein folding (Levinthal's paradox) and other NP-hard satisfiability (SAT) problems. As in these problems, we find that refolding a sheet requires actuation at multiple carefully chosen creases. We show that seeding successful folding in this way can be understood in terms of sub-patterns that fold when cut out (`folding islands'). Besides providing guidelines for the placement of active hinges in origami applications, our results point to fundamental limits on the programmability of energy landscapes in sheets. 
\end{abstract}

\maketitle

\section*{Introduction}

Single degree of freedom mechanical structures are attractive in a range of fields as almost any force will actuate that specific designed mode. Much like an umbrella or a folding chair, such `self-folding' structures can be reliably deployed even in uncertain environments with unreliable actuation forces. This principle has found wide use in kinetic or deployable architecture, heart stents, MEMS, sensors and robots on a range of length scales \cite{Peraza-Hernandez2014-gp,silverberg2014,Pellegrino:2014vg,Reis:2015ey}; recently, self-folding origami has become a popular framework for such applications \cite{Lang:2007,Demaine:2007vk,Tachi2009-kh,Tachi:2010tk,Na:2015wu,Dudte2016-ws}. 

The self-folding approach is similar in spirit to other bottom-up methods such as self-assembly of particles \cite{Hormoz2011-vg} and self-folding of polymers \cite{Pande2000-fg}; these methods exploit careful programming of interactions to allow for careless actuation at deployment. However, in these other self-actuating frameworks, the interactions needed for the desired assembly or folding inevitably create many other `distractor' states (e.g., kinetic traps in self-assembly \cite{Jacobs2015-jt,Zeravcic2014-yb,Murugan2015-wb} or in protein folding \cite{Pande2000-fg,Abkevich1994-bq,Karplus1997-sj}), necessitating more care at deployment than one would naively expect.

\begin{figure}	
\includegraphics[width=1\linewidth]{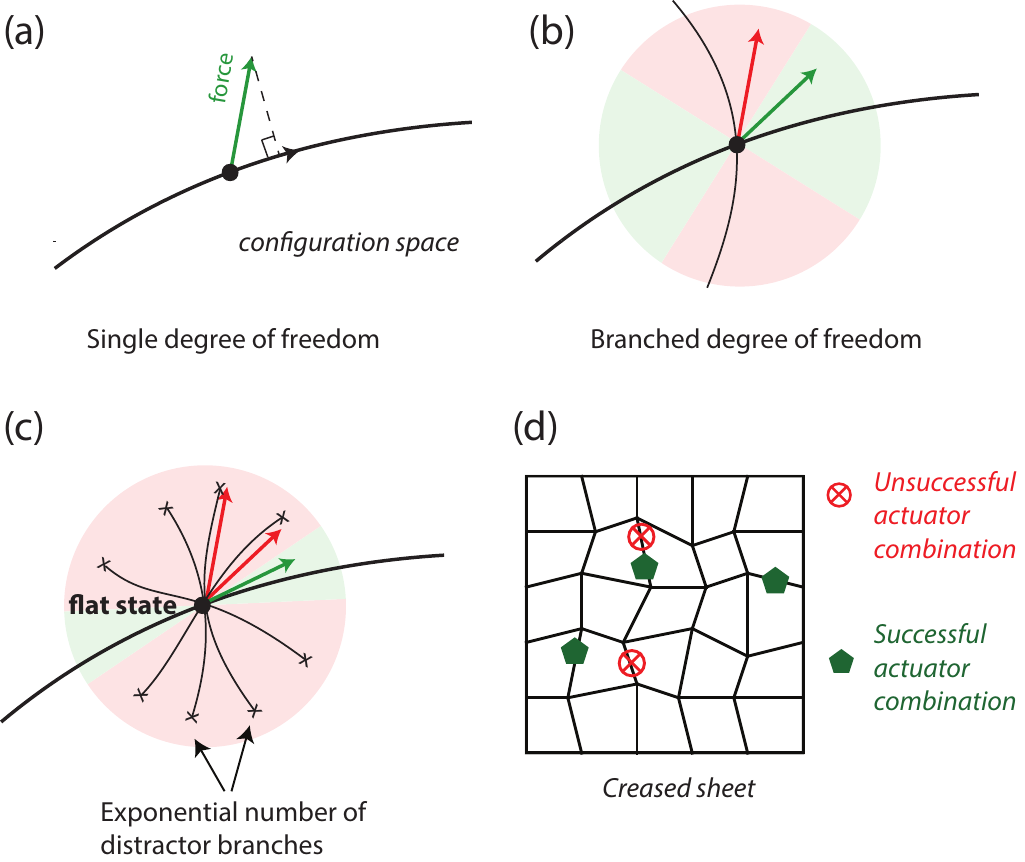}
\caption{(a) Structures designed with only one folding motion (`mechanisms') are thought to be easy to control since any applied force not exactly perpendicular to that motion will actuate it. (b) However, if a mechanism has a branched degree of freedom (bifurcation), the applied force (green) must make a smaller angle with the desired branch than with the undesired branch. (c) We show that programming a stiff sheet with one folding motion inevitably creates an exponential number of other dead-end `distractor' branches that are of zero energy only to linear order. The applied force needs to be highly aligned with the desired folding motion in order to avoid the distractors. (d) Consequently, we must actuate multiple creases in a carefully selected combination (green) to successfully fold a self-folding crease pattern.
\label{fig:distractors}}
\end{figure}

Here, we show that folding self-folding origami (a thin sheet pre-creased to allow only a single folding motion) is difficult because of a similar inevitable proliferation of distractor folding branches. The distractor branches, shown schematically in Fig.~\ref{fig:distractors}, meet at a bifurcation at the flat state but are dead-ends since they are of zero energy only to linear order. The number of distractors grows exponentially with the boundary length of the sheet and consequently, most spatial distributions of folding forces will actuate a distractor, as shown in Fig.~\ref{fig:distractors}c,d. As a result, despite having only one extended degree of freedom, self-folding crease patterns require multiple actuators placed at carefully chosen spatial locations for successful actuation.

We trace the origin of distractors to frustrated loops of vertices, each of which can fold along one of two branches. Such frustrated loops create a glassy energy landscape for the sheet around the flat state with an exponential number of local minima corresponding to the distractors. Successful folding must be seeded by actuation at a critical set of creases that picks out the ground state of the glassy landscape, much like with protein folding \cite{Levinthal1969-wm, Crescenzi1998-uq, Thomas_Ngo1994-xk, Karplus1997-sj} and other satisfiability problems \cite{Biroli2002-hf}. We find that the spatial arrangement of actuators needed can be understood heuristically in terms of unfrustrated `folding islands', the largest sub-pattern containing a given actuated crease that will fold when cut out of the full pattern.

Our results show the limits of programmability of energy landscapes for self-folding sheets, paralleling similar limitations due to undesired but inevitable traps in other bottom up approaches like self-assembly \cite{Hormoz2011-vg} and self-folding polymers \cite{Pande2000-fg}. Besides the theoretical significance, our results provide a practical means of understanding where to place active creases; e.g., in hydrogels or shape memory alloys, one must choose the active hinges; our theory predicts which combination of hinges would be successful and even predicts that sometimes, adding a new active crease (aiding in the right direction) to an existing successful actuation can in fact prevent folding. 

Our results on glassiness and the difficulty of physically folding origami superficially resembles earlier works, such as Bern and Hayes' classic result on NP-hardness of flat-foldability \cite{Bern1996-pn} and others \cite{Abel2015-nf, Ballinger2015-rk,Arkin2000-tb, Demaine2007-td}. 
However, Bern and Hayes focused on the ordering of folds in multi-stage folding, also investigated later in \cite{Pandey2011-cc, Hawkes2010-qr, An2011-um}. Here, we focus on self-folding sheets with a single temporal stage. More critically, many earlier works \cite{Bern1996-pn, Arkin2000-tb} concern the \textit{computational} difficulty in finding a consistent global Mountain-Valley assignments (e.g., `forcing sets' \cite{Abel2015-nf,Ballinger2015-rk}), while our work concerns whether the \textit{physics} of folding can find a desired global Mountain-Valley assignment, taking into account physical effects such as mechanical advantage and energy landscapes that play no role in these earlier works. A recent  work \cite{Tachi2016-si} considers similar actuation questions for single vertices and quads; in contrast, we use an energy model and focus on statistical results for large quad meshes with an exponential number of distractors.

\begin{figure}	
\includegraphics[width=1\linewidth]{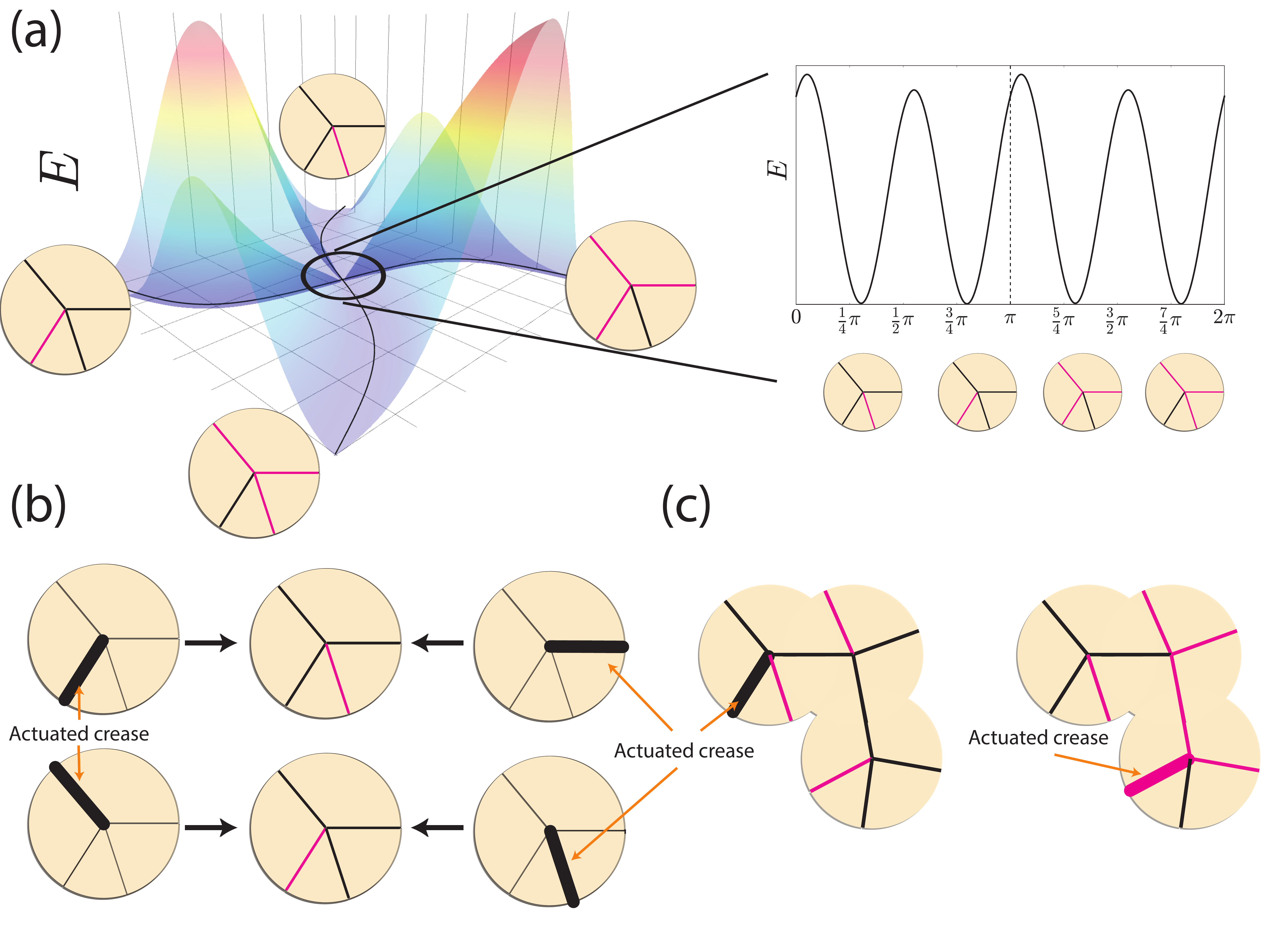}
	\caption{Bifurcations for vertices and chains of vertices. (a) A single vertex has two distinct folding branches that meet at a bifurcation at the flat state. The Mountain-Valley pattern of the two branches differ in the placement of their `odd-one-out' crease (e.g., the single Valley crease in a branch with 3 Mountains). (b) When a selected crease is actuated, the vertex chooses the branch in which that actuated crease folds more relative to other creases (rule of mechanical advantage). Since the odd-one-out crease and its transverse crease tend to fold less than the other crease pair, the odd-one-out crease is generally adjacent to the actuated crease. (c) When $N$ vertices are linked together into an open-ended chain, the chain can fold in $2^N $ different folding branches. Given an actuated crease, the resulting MV data can be predicted by applying the branch selection rule of (b) to vertices in sequence, as each successive vertex is actuated through the crease linking it to the prior vertex. \label{fig:vertices}}
\end{figure}

\begin{figure}	
\includegraphics[width=1\linewidth]{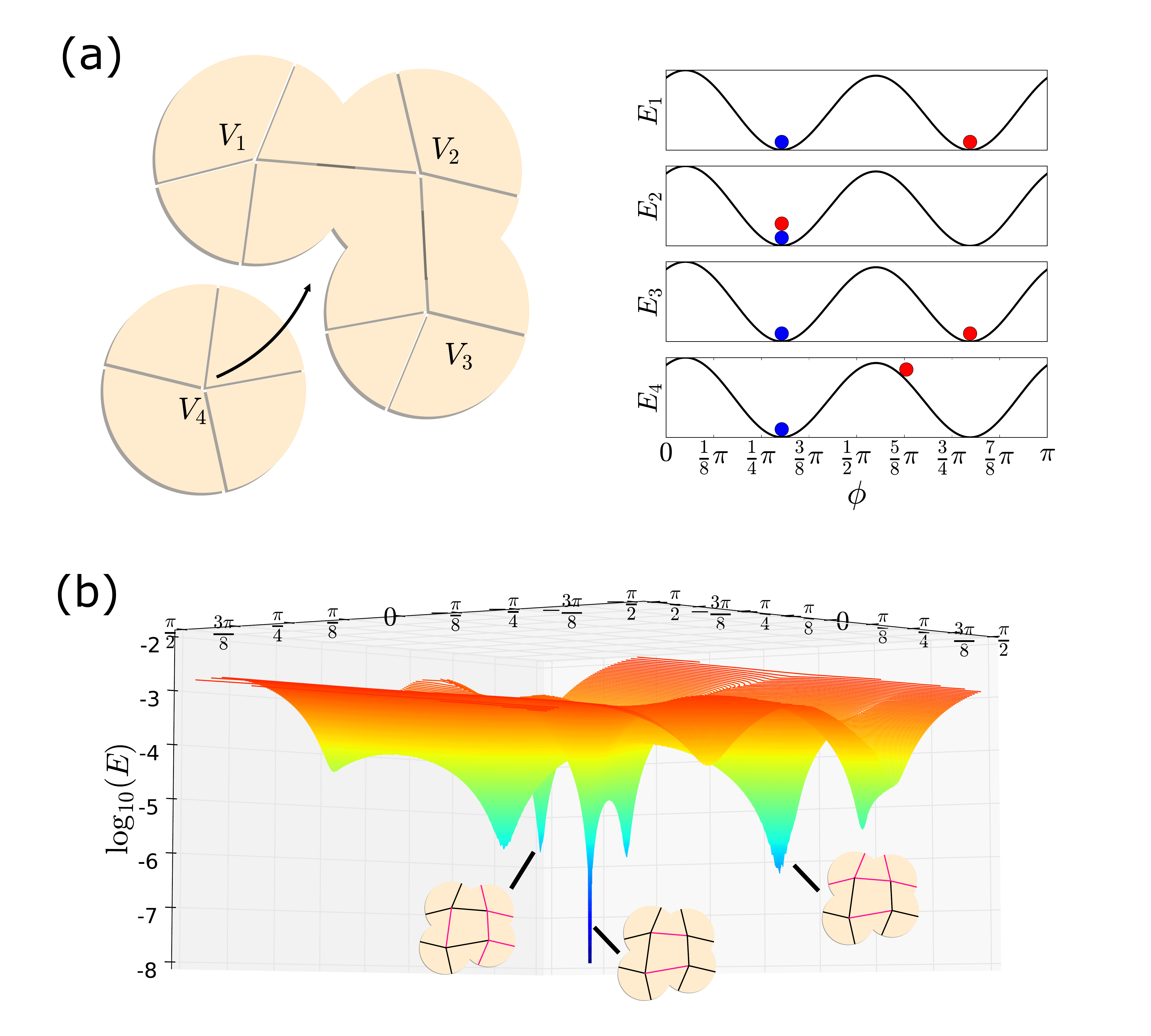}
	\caption{A loop of vertices gives rise to a glassy landscape. (a) When a chain of $N=3$ vertices is closed by adding a fourth vertex, the resulting $2^3$ branches are no longer of zero energy. For example, if vertices $V_1,V_2,V_3$ are at one of their two zero energy states (red dots), $V_4$'s folding state is completely determined because the folding state of creases $V_{1}-V_{4}$ and $V_{3} - V_{4}$ are already set. The resulting energy of $V_4$ may not be zero (red dot for $E_4$). (b) To visualize the energy landscape of the pattern as a whole, we went through all possible folding angles within the linearized null-space; we find a glassy energy landscape for the crease pattern at fixed norm $||\rho||$ of the crease folding angles $\vec{\rho}$ (a 2-dim projection is shown). One select minimum is of zero (or particularly low) energy if the geometry of the crease pattern had been designed to exhibit such a folding branch. Successful folding in the desired branch requires initiating folding in the narrow attractor basin of that minimum. \label{fig:quad}}
\end{figure}

\section*{Results}
\subsection*{4-vertex and chains of 4-vertices}
When a vertex with $n$ creases is folded, the $n$ dihedral fold angles $\rho_i, i = 1,\ldots ,n$ are related by $3$ equations \cite{Huffman1976-rr}. Thus $n$-valent vertices with $n\leq 3$ will be completely rigid, while vertices with $n\geq 5$ have multiple degrees of freedom. 4-vertices are of special interest as they have precisely one degree of freedom. 

However, a crucial caveat to this Maxwell counting is that only $2$ of the $3$ vertex equations are independent when the vertex is laid out flat \cite{Tachi2016-si} (i.e., unfolded). Consequently, it was shown \cite{Waitukaitis2015-rw} that a generic $4$-vertex has two distinct folding branches that meet at a bifurcation at the flat state. See Fig.~\ref{fig:vertices}. 

To see this quantitatively, we follow Tachi's use of rotation matrices \cite{Tachi2010-qg,hull:2002} to derive three constraint equations $T_a(\vec{\rho},\vec{\theta}) = 0, a= 1,2,3$ associated with the vertex where $\vec{\rho}$ are the fold angles at creases and $\vec{\theta}$ are the in-plane angles between creases (see Supplementary Information). We expand the constraints $T_a$ in a series in $\rho_i$ about the flat state $\vec{\rho} = 0$ as $T_a(\vec{\rho}) = C_{a}^i \rho_i + D_{a}^{ij} \rho_i \rho_j + \ldots$ (where repeated indices are summed over). If these constraints are violated by a configuration, so $T_a(\vec{\rho}) \neq 0$, the faces of the vertex will bend to accommodate the required crease folding; we can associate an energy,
\begin{equation}
E = \sum_a T_a^2 = \sum_a ( C_{a}^i \rho_i + D_{a}^{ij} \rho_i \rho_j + \ldots)^2
\label{eqn:energy}
\end{equation}
with these face-bent configurations (see Supplementary Information for more details on the energy model).

The energy of a general configuration (Eq.~\ref{eqn:energy}) scales as $||\rho||^2$. However, $C_{a}^i$ has rank 2, giving a two dimensional space of zero modes in the linear approximation about the flat state $\vec{\rho}=0$. The energy scales as $||\rho||^4$ for folding modes in this linearized null space. Fig.~\ref{fig:vertices}a shows the energy for folding modes within the linearized null space as we fold to larger angles. We see that two special folding branches within the linearized null space have zero energy to all orders. Thus, a generic $4$-vertex has a full $2d$ vector space of zero modes at the flat state in a linear approximation, but only two $1d$ branches of zero energy upon non-zero folding. This is consistent with Maxwell counting, as one constraint is redundant at, but only at, the flat state.

The two folding branches differ qualitatively in the sign of their fold angles. Both branches satisfy the following rule \cite{kawasaki1989,hull1994}; three of the four creases must fold in a common orientation (say, `Valley' fold) with the final `odd-one-out' crease folding the other way (`Mountain' fold). The final odd-one-out crease can be either one of the two creases whose neighboring angles add to less than $\pi$; see Fig.\ref{fig:vertices}a. This discrete choice gives rise to the two branches. Note that the two creases capable of being the odd-one-out are always adjacent.

\subsection*{Branch selection through mechanical advantage}

When external folding torques $\tau_i, 1=1\ldots 4$ are applied to the creases of a 4-vertex and released, the vertex will relax into one of the two branches (see Fig.~\ref{fig:vertices}a) with corresponding folding angles $\vec{\rho}_\alpha,\alpha = 1,2$. In the linear regime $||\vec{\rho}||\ll 1$, using our model of energy in Eqn.\ref{eqn:energy}, we find that computing the normalized dot product between the applied vector of torques (henceforth `applied force') $\vec{\tau}$ and the folding angles $\vec{\rho}_\alpha$ of the two branches identifies the actuated branch; the vertex will relax into the branch with higher dot product $\tau \cdot \vec{\rho}_\alpha / ||\vec{\tau}|| ||\vec{\rho}_\alpha||$. This rule is equivalent to selection based on mechanical advantage; when only one crease is actuated, the vertex folds into the branch in which that crease's folding is larger relative to other creases (i.e., contributes more to the norm $||\vec{\rho}_\alpha||$). 

Our mechanical advantage rule is based on a model energy landscape where the angular bisector of the two branches separates their attractor basins. In real material vertices, the dividing line between the attractors might be closer to one branch than the other; such complications do not change our results qualitatively. In contrast, a recent work \cite{Tachi2016-si} assumed that actuation might fail if the applied force has a positive dot product with \textit{any} other available branch. In such a model, even applied forces perfectly aligned with a branch may be classified as incapable of evoking that branch, in contrast to energy landscape-based models.

Our mechanical advantage rule can be restated as a heuristic in terms of Mountain-Valley (MV) choices. In either folding branch, the crease with odd-one-out MV state and its transverse crease fold less than the other pair of creases that share a common MV state. (To see this intuitively, consider the limiting case in which all in-plane angles are nearly $90$ degrees and the vertex simply folds in half along one pair of creases with the same MV state; the other pair of creases barely fold at all.) 

Combining this observation with the dot product rule, we conclude that when a single crease is actuated, the vertex will choose the branch in which the crease transverse to the control crease will fold with the same MV state. See Fig.~\ref{fig:vertices}b. 

This branch-picking rule is easily extended to chains or trees of vertices, as long as no loops are present. If we actuate at one select crease at a vertex in this chain, we can determine the branch choice at that vertex using the above rule and thus the MV state of all creases at that vertex. Any neighboring vertex is actuated by the creases connecting them. In the absence of loops, there is only one path from the controlled vertex to any other and hence the mode-propagation rule unambiguously determines the branch choice at each vertex.

In this way, for any given actuated crease, the branch selection and propagation rule unambiguously selects one branch out of the $2^N$ bifurcated folding branches of an $N$ vertex chain.

\subsection*{Loops of vertices create glassy energy landscapes}

If 4-vertices are connected around a loop, we can no longer make an independent choice of folding branch at each of the vertices. For example, for a loop of four 4-vertices like that in Fig.~\ref{fig:quad}a, we can make independent branch choices for three of the vertices - say for $V_1,V_2$ and $V_3$ - which puts them in one of their zero energy states (red or blue points in Fig.~\ref{fig:quad}a). The final vertex's folding state is then completely determined because the state of two creases at $V_4$ are already determined (namely, creases $V_3-V_4$ and $V_4-V_1$). Generically, the resulting state for $V_4$ will not be of zero energy \cite{Silverberg:2015gb} (red dots in Fig.~\ref{fig:quad}a). We thus find that the resulting folding branch is of non-zero energy, unlike for chains of vertices.

Going through the $2^3 = 8$ independent branch choices for $V_1,V_2,V_3$ (which then determine the state of $V_4$), we should expect to generically find $8$ branches of non-zero energy. In fact, these folding branches are of zero energy to quadratic order but of non-zero energy at next order; i.e., the energy of these branches scale as $\kappa_i \rho^4$ with $\kappa_i \neq 0$. In contrast, $\kappa_i = 0$ for all the $2^N$ folding branches of a chain of vertices. 

To gain more intuition about these branches and their energies $\kappa_i \rho^4$, we fixed the overall folding magnitude $||\vec{\rho}||$ for a single quad and computed the energy ($\propto \kappa_i$) as a function of the angular directions in $\vec{\rho}$ space. A two-dimensional projection is shown in Fig.~\ref{fig:quad}b where each branch shows up as a local minimum with depth proportional to $\kappa_i$.

Thus, we find that loops of vertices have a glassy folding energy landscape, much like a spin network with frustrated loops \cite{binder1986}, and unlike trees or chains of spins. 

A desired branch's energy can be made arbitrarily low or even zero to all orders in folding by fine-tuning in-plane angles using `loop' equations \cite{Tachi:2012,OurOwnPaper}. While the design process can make a desired folding branch be the ground state of the landscape, it does not change the glassy attractor structure shown in Fig.~\ref{fig:quad}b; see SI for a comparison. Different actuated creases initialize the folding process in different parts of the glassy landscape; folding then involves flowing downhill to a local minimum. Hence, actuating a desired branch in such a landscape can be difficult in the presence of a multitude of distractor branches.

\subsection*{Large patterns - number, attractor size of distractors}
\begin{figure}	
\includegraphics[width=1\linewidth]{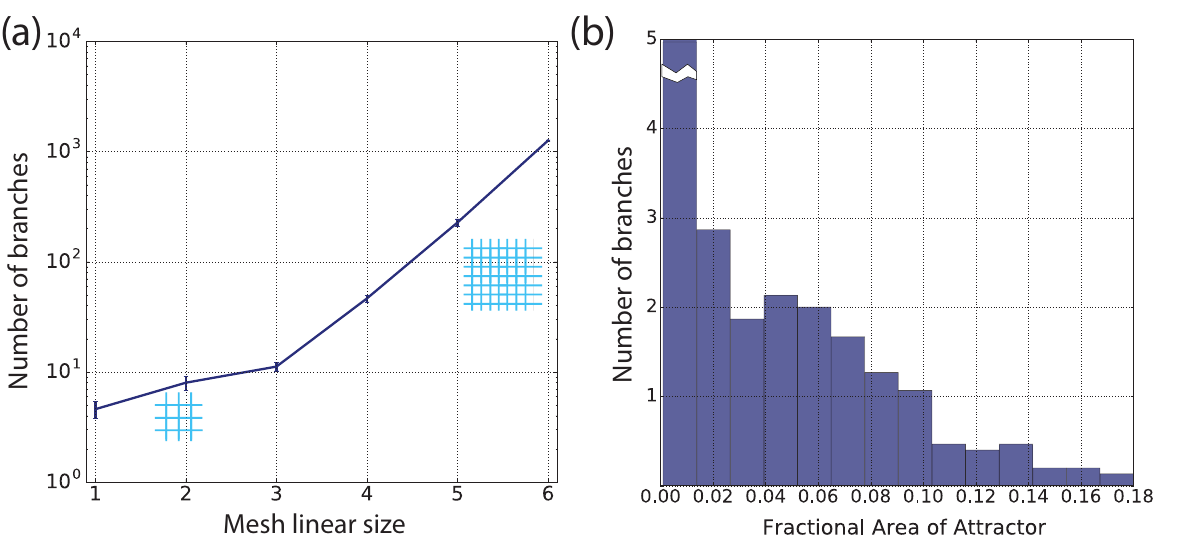}
\caption{Large patterns have an exponential number of branches (i.e., minima) of decreasing attractor size. We characterized the landscape by sampling random quad meshes of size up to $A = 36$ and folded each one with $\sim 10^3$ random torques. (a) A quad mesh of size $\sqrt{A} \times \sqrt{A}$ quads has $\sim 2^{\alpha \sqrt{A}}$ distinct local minima (i.e., folding branches) in its energy landscape ($\alpha \approx 2.25$). (b) The size of attractor basins around different branches for a fixed pattern does not exceed $17\%$ of the total space for a $4 \times 4$ mesh.  
\label{fig:attractors}}
\end{figure}

Large patterns made of many 4-vertices contain many loops and the number of distractor branches grows rapidly. We generated quad meshes of random geometry made of $\sqrt{A} \times \sqrt{A}$ quadrilateral units, folded each quad mesh with $2000$ random applied forces $\vec{\tau}$ and allowed it to relax into a local energy minimum. In this way, we determined the following landscape properties:

(a) total number of distinct branches $N_{branches}$ for a given quad mesh grows as 
\begin{equation}
N_{branches} \sim 2^{\alpha \sqrt{A}}
\end{equation}
with $\alpha \approx 2.25$; see Fig.~\ref{fig:attractors}a. Why does the number of distractor branches scale only as $2^{\alpha \sqrt{A}}$ and not $2^{\alpha A}$, given that each of the $O(A)$ vertices has two distinct choices of a branch? The reduction is due to loop behavior shown in Fig.~\ref{fig:quad}. Consider making a independent choice of branch for all of the boundary vertices. As shown in Fig.~\ref{fig:quad}, once three vertices in a quad have been set, the folding state of the fourth vertex is completely determined. We can iterate this argument to determine the folding state of all the bulk vertices for any independent choice of branches along the boundary.

(b) the attractor size of each distractor branch (taken to be the fraction of random actuation forces that actuate the branch) is generally small; see Fig.~\ref{fig:attractors}b. The largest attractor for the $4 \times 4$ mesh sampled is only $\approx 17\%$ i.e., only $17\%$ of random torques will actuate that branch. Most branches have far smaller attractor basins (Modes with attractor size smaller than the sampling error of $\sim 2\%$ are not shown).

The typical attractor size is expected to drop exponentially with $A$ as $2^{-\alpha \sqrt{A}}$ since the number of attractors grows as $2^{+\alpha \sqrt{A}}$. Simulations of a $6 \times 6$ lattice suggest that no mode has an attractor size exceeding $0.2 \%$.

\subsection*{Actuation of large loopy patterns}

How many creases need to be actuated - and which ones -  to pick the desired branch in a glassy landscape with an exponential number of other minima?

\begin{figure*}	
\includegraphics[width=1\linewidth]{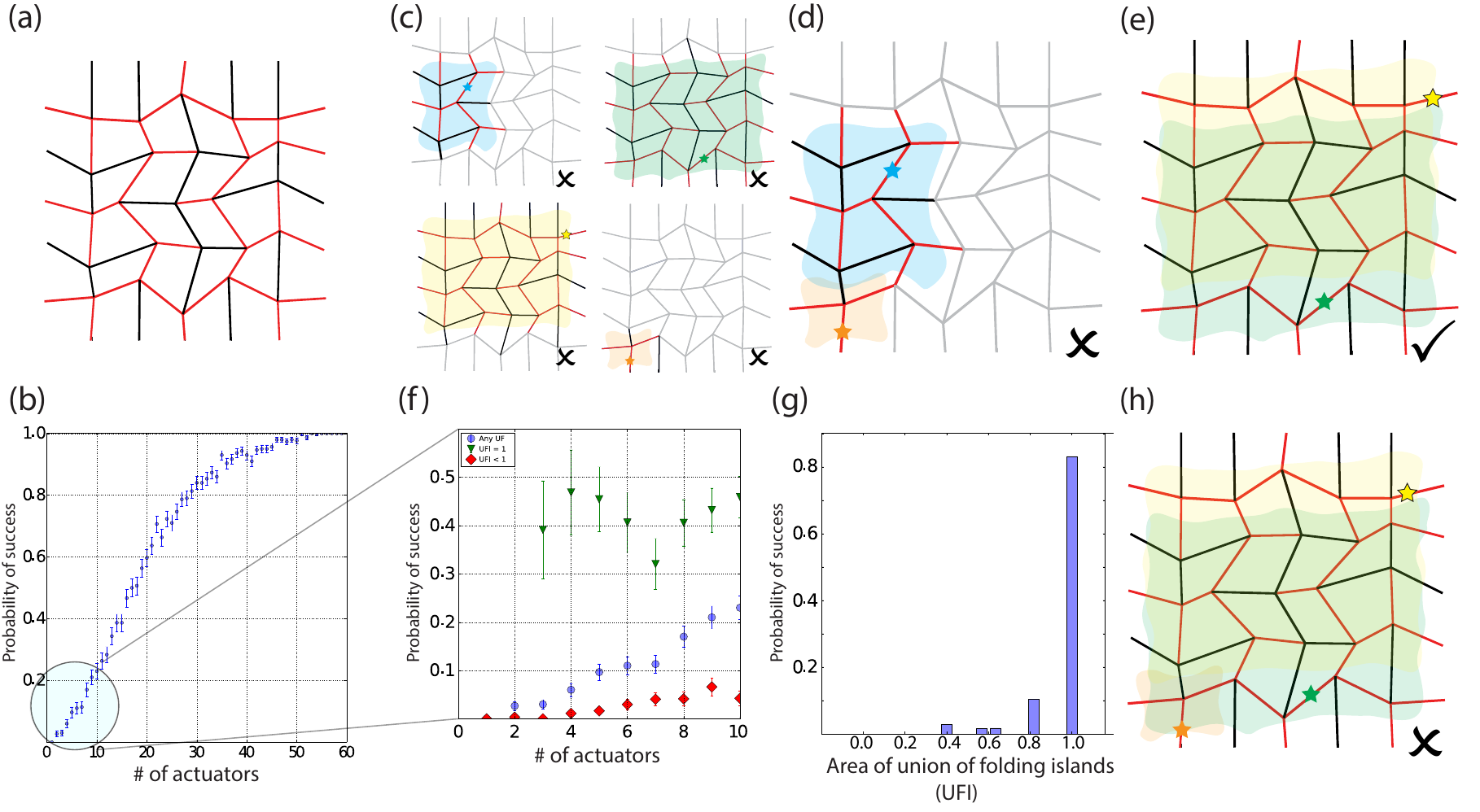}
	\caption{Spatial distribution of acutators determines folding success. If actuators are placed on randomly chosen creases of the pattern shown in (a), at least $18$ actuators ($\sim 30\%$ of creases) are needed to have a $50\%$ chance of successful folding, as shown in (b). The `folding island' of a crease, shown for four typical creases in (c), is the largest sub-pattern that will fold correctly when cut out from the full pattern and actuated at that crease. (d,e) The area of the union of folding islands relative to the entire pattern (denoted $UFI$) provides a simple design heuristic; (d) Actuated crease sets with $UFI<1$  generally do not successfully fold the pattern while the actuators in (e) with $UFI=1$ are successful. (f) Actuated crease sets of given size are dramatically more likely to fold successfully if their $UFI = 1$ (green) rather than $UFI < 1$ (red). (g) Replotting the data from (b) for all crease sets, we find $UFI$ is a sharper predictor of success than the number of actuators. (h) Folding islands also explain a counter-intuitive effect where successful actuation, as shown in (e), can sometimes be ruined by actuating an additional crease (orange) with a small folding island. \label{fig:foldingislands}}
\end{figure*}

To answer this question, we study a random pattern with a chosen branch, shown in Fig.~\ref{fig:foldingislands}a. Since the crease locations at which folding torques are applied can be better controlled than the precise magnitude of torque \cite{Peraza-Hernandez2014-gp} in many applications, we applied folding torques of fixed $O(1)$ magnitude to different randomly selected subsets of creases. The applied torques were always of the correct sign (Mountain or Valley) needed at that crease for the chosen branch. As seen in Fig.~\ref{fig:foldingislands}b, actuators are needed on $18$ out of a total of $60$ creases to even have a $50\%$ probability of folding the pattern. 

For applications where the precise torque magnitudes can be controlled in addition to location (as explored recently in \cite{Tachi2016-si}), we must characterize how closely the applied vector of torques must align with the folding angles of the desired branch (see Fig.\ref{fig:distractors}). We present such results on dot products in the SI. 

Requiring a large number of actuators or precise control of torque magnitudes defeats the purpose of designing a single degree of freedom mechanism; it is hard to call a system requiring such delicate control `self-folding'. 

How then can self-folding origami be folded with a minimal number of actuators? A lesson can be drawn from similar glassy landscape search problems in models of protein folding (e.g., Levinthal's paradox \cite{Thomas_Ngo1994-xk, Karplus1997-sj,Berger1998-ar, Crescenzi1998-uq}) and related NP-hard satisfiability (SAT) problems \cite{Biroli2002-hf,Krzakala2008-nh} that vary from the Traveling Salesman Problem to Sudoku \cite{Ines_Lynce2006-eo}. A common element in these satisfiability problems is that random seeding of the search for the global minimum leads to repeated backtracking after reaching local minima, both in the context of computer algorithms (as the DPLL algorithm for k-SAT \cite{Biroli2002-hf}) or for physical dynamics (as in protein folding) \cite{Krzakala2008-nh}. However, careful seeding of the search - e.g., if the right boxes are filled in first in Sudoku \cite{Ines_Lynce2006-eo} or if the right parts of the protein are folded first - can greatly reduce or even eliminate backtracking \cite{Biroli2002-hf} before reaching the global minimum. 

Correct seeding is even more critical for origami since folding is assumed to happen at `zero temperature' (e.g., without any noise or fluctuations).  As a result, the structure cannot backtrack out of a local minimum as in the case of non-zero temperature SAT problems \cite{Krzakala2008-nh}.

\subsubsection*{Folding islands}

To understand the role of frustration and seeding in the origami context, note that the branch selection rule, illustrated for vertex chains in Fig.~\ref{fig:vertices}b, can be ambiguous when applied to loops. In the presence of loops, the MV state can be propagated from a control crease to a target vertex along multiple different paths using our mechanical advantage rule which acts to ensure that the crease from which folding has propagated and the crease transverse to it fold with the same MV state. The propagated state along different paths may not agree with each other - and critically - can disagree with the desired folding branch for the target vertex. In such a case, the resulting vertex might fold incorrectly. Thus, while a designed folding branch guarantees a \textit{globally} consistent configuration of vertex branch choices (e.g., blue dots in Fig.\ref{fig:quad}), such a global configuration may be difficult to reach using the \textit{local} MV propagation rule in Fig.\ref{fig:vertices} from to a single actuator. Hence successfully folding along a branch for large patterns can require actuating multiple creases at the same time. 

To find the number of actuators needed, we identify unfrustrated sub-patterns called `folding islands'. We define the folding island of a crease (with respect to a desired folding branch) as the largest subset of the pattern that will fold in the desired branch, if that subset is cut out and actuated at the chosen crease. Fig.~\ref{fig:foldingislands}c shows that folding islands for different creases can vary greatly and generally do not cover the whole pattern. While folding islands can be approximately deduced using the simple MV propagation rule in Fig.~\ref{fig:vertices}b, the exact shape can depend on the precise in-plane angles.

These considerations suggest a heuristic necessary condition for a set of actuated creases to fold a pattern; the union of their folding islands should cover the whole pattern. If not, as in Fig.~\ref{fig:foldingislands}d, when folding reaches the boundary of a folding island, folding will jam in one of the high energy distractor branches because vertices just outside the union of islands will fold incorrectly. On the other hand, the two actuators shown in Fig.~\ref{fig:foldingislands}e, whose folding islands together cover the entire pattern, successfully fold the pattern.

Folding islands provide a new perspective on why randomly placed actuators (Fig.~\ref{fig:foldingislands}b) were poor at folding the pattern. In Fig.~\ref{fig:foldingislands}f, we went through the different actuated crease sets used in Fig.~\ref{fig:foldingislands}b and computed the area of the Union of Folding Islands (which we denote $UFI$, defined as the fraction of all creases belonging to the union) for each set. We see, for example, that a set of 5 actuators is $60\times$ more likely to fold the pattern if it has $UFI=1$ rather than $UFI < 1$. Similarly, Fig.~\ref{fig:foldingislands}g shows all the data in Fig.~\ref{fig:foldingislands}b, but plotted against $UFI$ instead of number of actuators. Together, these results show that the union of folding islands and thus spatial placement of actuators is a much better predictor of folding success than just the number of actuators. (We do find a few cases of successful actuation e.g., at $UFI = 0.8$ when the folding islands cover all but a few boundary vertices). In particular, the condition $UFI = 1$ eliminates many spatial arrangements of actuators that are nearly guaranteed to fail.

Folding islands also shed light on a counterintuitive phenomenon shown in Fig.~\ref{fig:foldingislands}h. While the two actuators in Fig.~\ref{fig:foldingislands}e can successfully fold the pattern, adding another actuator with a very small folding island as in Fig.~\ref{fig:foldingislands}h, can stop the previously successful folding! In this case, the vertices just outside the island of the orange crease are folded incorrectly. (Such effects reduce the probability of success in Fig.~\ref{fig:foldingislands}g when $UFI = 1$ to be less than $1$). Predicting such subtle competition between the different control creases requires knowledge of the precise in-plane angles of the pattern and we are unable to formulate a strict necessary and sufficient condition for successful folding without full pattern information. Nevertheless, identifying the folding islands provides a useful design heuristic to greatly reduce the number of actuators needed, as seen in Fig.~\ref{fig:foldingislands}f,g.

\section*{Discussion}

We showed that sheets with crease patterns designed to exhibit exactly one folding behavior are nevertheless difficult to fold. We traced this difficulty to the fact that stabilizing one folding behavior using frustrated interactions between binary degrees of freedom (bifurcated origami vertices \cite{Waitukaitis2015-rw,Huffman1976-rr}) inevitably stabilizes an exponential number of other distractor behaviors. Thus our results establish fundamental limits on the programmability of energy landscapes for sheets, paralleling similar limitations in other bottom-up approaches such as self-assembly of particles \cite{Hormoz2011-vg} and self-folding of polymers \cite{Pande2000-fg} as well as classic NP-hard satisfiability (SAT) problems \cite{Krzakala2008-nh,Biroli2002-hf}.

We saw that many actuators are needed to successfully fold self-folding sheets, if their locations are randomly chosen. However, carefully choosing the set of actuated creases can reduce their number dramatically; we interpreted such successful combinations in terms of unfrustrated sub-patterns called folding islands that successfully fold when cut out of the full pattern.

Recent self-folding origami applications vary greatly in the materials used and in actuation mechanisms for active hinges, including electric \cite{Peraza-Hernandez2014-gp}, optical \cite{Zanardi_Ocampo2003-oj}, thermal \cite{Kuribayashi2006-rs} and chemical (pH) \cite{Shim2012-jj} methods. In many applications, energy can be selectively input to specific creases, e.g., by controlling the electric current to shape-memory polymer hinges \cite{Felton2013-kf,Hawkes2010-qr} or light input to hydrogels \cite{Silverberg2015-ut}. Our work suggest which combinations of creases should be given energy input for successful folding, even showing how adding an actuator can ruin successful folding (Fig.~\ref{fig:foldingislands}h). 
Going beyond self-folding patterns, our considerations also apply to each temporal stage of multi-stage sequential folding patterns \cite{Hawkes2010-qr,An2011-um,Pandey2011-cc}.

The folding difficulty described here and the resulting need for careful actuation mathematically applies only at the flat state; but since the energy barriers between branches grow more slowly with folding for a softer sheet, careful actuation needs to be maintained until a larger folding angle for soft sheets.

Recent experiments on controlled repeated crumpling and extension of sheets suggests an inability to refold along existing creases, leading to the formation of new creases \cite{gottesman}. While the $4$-vertex patterns studied here are not good models of crumpled soft paper with significant face bending, our results do suggest that the difficulty of refolding a crease pattern, and thus the propensity to create new creases, grows with the softness of the sheet and when unfolded closer to the flat state.

\begin{acknowledgments}
We thank William Irvine, William Jacobs, Yoav Kallus, Christian Santangelo, Cristopher Moore, and Thomas Witten for insightful discussions. We acknowledge NSF-MRSEC 1420709 for funding and the University of Chicago Research Computing Center for computing resources.
\end{acknowledgments}

\appendix

\bibliographystyle{unsrt}
\bibliography{Paperpile_-_Mar_10_BibTeX_Export,NachiCitations}

\end{document}